\documentclass[%
 aip,
 amsmath,amssymb,
 reprint,%
]{revtex4-1}

\usepackage{graphicx}
\usepackage{dcolumn}
\usepackage{bm}

\usepackage[utf8]{inputenc}
\usepackage[T1]{fontenc}
\usepackage{mathptmx}
\usepackage{etoolbox}
\usepackage{textcomp}

\makeatletter
\def\@email#1#2{%
 \endgroup
 \patchcmd{\titleblock@produce}
  {\frontmatter@RRAPformat}
  {\frontmatter@RRAPformat{\produce@RRAP{*#1\href{mailto:#2}{#2}}}\frontmatter@RRAPformat}
  {}{}
}%
\makeatother
\begin{document}

\preprint{AIP/123-QED}

\title[]{Cryogenic enhancement of phononic four-wave mixing in AlScN/SiC}

\author{A. K. Behera}
\thanks{These authors contributed equally.}
\affiliation{Center for Integrated Nanotechnologies, Sandia National Laboratories, Albuquerque, New Mexico 87185,
USA}

\author{B. Smith}
\thanks{These authors contributed equally.}
\affiliation{Microsystems Engineering, Science, and Applications, Sandia National Laboratories, Albuquerque, New Mexico 87123, USA}

\author{X. Du}
\affiliation{Department of Electrical and Systems Engineering, University of Pennsylvania, Philadelphia, Pennsylvania 19104, USA}

\author{Y. Deng}
\affiliation{Electrical, Computer and Energy Engineering, University of Colorado, Boulder, Colorado 80309, USA}

\author{M. Miller}
\affiliation{Microsystems Engineering, Science, and Applications, Sandia National Laboratories, Albuquerque, New Mexico 87123, USA}

\author{N. Sagartz}
\affiliation{Microsystems Engineering, Science, and Applications, Sandia National Laboratories, Albuquerque, New Mexico 87123, USA}

\author{M. Koppa}
\affiliation{Microsystems Engineering, Science, and Applications, Sandia National Laboratories, Albuquerque, New Mexico 87123, USA}

\author{C. T. Harris}
\affiliation{Center for Integrated Nanotechnologies, Sandia National Laboratories, Albuquerque, New Mexico 87185,
USA}

\author{M. Lilly}
\affiliation{Center for Integrated Nanotechnologies, Sandia National Laboratories, Albuquerque, New Mexico 87185,
USA}

\author{R. H. Olsson III}
\affiliation{Department of Electrical and Systems Engineering, University of Pennsylvania, Philadelphia, Pennsylvania 19104, USA}

\author{M. Eichenfield}
\affiliation{Center for Integrated Nanotechnologies, Sandia National Laboratories, Albuquerque, New Mexico 87185,
USA}
\affiliation{Microsystems Engineering, Science, and Applications, Sandia National Laboratories, Albuquerque, New Mexico 87123, USA}
\affiliation{Electrical, Computer and Energy Engineering, University of Colorado, Boulder, Colorado 80309, USA}

\author{L. Hackett*}
 \email{lahacke@sandia.gov}
\affiliation{Microsystems Engineering, Science, and Applications, Sandia National Laboratories, Albuquerque, New Mexico 87123, USA}

\date{\today}

\begin{abstract}
Surface acoustic wave platforms based on piezoelectric thin-film heterostructures provide sub-wavelength acoustic confinement, making them attractive for compact nonlinear phononic systems with applications including frequency conversion, parametric interactions, and nonlinear signal processing. Here, we investigate guided surface acoustic wave phononic four-wave mixing at gigahertz frequencies in an aluminum scandium nitride (Al$_{0.58}$Sc$_{0.42}$N)/4H–silicon carbide heterostructure operated at both room temperature (295 K) and cryogenic temperature (4 K). The 500 nm thick aluminum scandium nitride film supports guided Rayleigh and Sezawa modes with distinct displacement and strain energy density distributions, allowing a direct comparison of mode-dependent nonlinear behavior within the same device. Continuous-wave four-wave mixing measurements reveal an enhancement in the extracted modal nonlinear coefficient at 4 K relative to 295 K for both modes. In addition, the Rayleigh mode exhibits a modal nonlinearity approximately two orders of magnitude larger than that of the Sezawa mode across both temperature regimes. These results demonstrate that phononic four-wave mixing is strongly influenced by temperature, mode confinement, and strain localization while establishing aluminum scandium nitride on silicon carbide heterostructures as a promising platform for engineering enhanced nonlinear phononic interactions for future classical and quantum acoustic on-chip signal processing systems.
\end{abstract}

\maketitle

The optical Kerr effect, arising from a material's third-order nonlinear susceptibility, $\chi^{(3)}$, has an acoustic analogue in which strain-dependent nonlinearities produce four-wave mixing between phonons.\cite{kurosu_mechanical_2020, mayor_gigahertz_2021, hackett_giant_2024} In photonics, Kerr nonlinearities underpin a wide range of classical and quantum technologies, including wavelength conversion, frequency-comb generation, and parametric amplification.\cite{lin_kerr_2022,stone_wavelength-accurate_2024,miller_-chip_2014,moody_chip-scale_2020} Establishing mechanisms to harness and optimize on-chip phononic third-order nonlinear processes could bring similar functionality into the acoustic domain for classical radio frequency (RF) signal processing,\cite{campbell_surface_2012} quantum phononics,\cite{dumur_quantum_2021, chou_deterministic_2025} and hybrid quantum acoustic platforms that interface phonons with superconducting or solid-state quantum systems.\cite{clerk_hybrid_2020, gustafsson_propagating_2014}

Surface acoustic waves (SAWs) confine acoustic energy within approximately one wavelength of the surface, resulting in large local strain in a small effective mode volume.\cite{campbell_surface_1998, royer_elastic_1999} In layered piezoelectric heterostructures, such as aluminum scandium nitride (AlScN) on silicon carbide (SiC), acoustic velocity mismatch between the thin film and substrate gives rise to guided slab-mode SAW phononic waveguides that exhibit enhanced vertical confinement over conventional SAWs. SAW-based technologies include chip-scale components for the RF front end such as high performance filters,\cite{lu_recent_2025} amplifiers,\cite{mansoorzare_micromachined_2022, hackett_non-reciprocal_2023, ghosh_acoustoelectric_2019} and oscillators,\cite{chang_voltage-controlled_2024, zhang_surface_2020, tanaka_lithium-niobate-based_2012, xi_low-phase-noise_2025, wendt_electrically_2025} as well as emerging quantum applications.\cite{pirkkalainen_hybrid_2013,maity_coherent_2020,arrangoiz-arriola_resolving_2019,maity_mechanical_2022} Further understanding mechanisms for leveraging nonlinearities in guided SAW modes provides a route to enhancing the functionality of existing SAW-based platforms. At the same time, a detailed understanding of phononic nonlinearities is essential for applications that demand highly linear operation, where suppressing unwanted nonlinear mixing is critical for preserving signal fidelity and quantum coherence.

On-chip phononic four-wave mixing has been previously demonstrated across several material platforms. A mechanical analogue of Kerr fiber optics was realized in a flexural-mode GaAs/AlGaAs nanoelectromechanical waveguide, where self-phase modulation, cross-phase modulation, and four-wave mixing were observed at megahertz frequencies.\cite{kurosu_mechanical_2020} Gigahertz-frequency four-wave mixing and cascaded sideband generation were shown in thin-film lithium niobate phononic integrated circuits.\cite{mayor_gigahertz_2021} In addition, electron-mediated phononic nonlinearities in semiconductor-piezoelectric heterostructures have been shown to significantly enhance the phononic third-order nonlinearity and thus the four-wave mixing power conversion efficiency compared to the case where the nonlinearity arises solely from the piezoelectric material.\cite{hackett_giant_2024,mansoorzare_acoustoelectric-driven_2023}

In this work, we study phononic four-wave mixing in a 500 nm thick aluminum scandium nitride/4H-silicon carbide (Al$_{0.58}$Sc$_{0.42}$N/SiC) thin film heterostructure at gigahertz frequencies. We carry out four-wave-mixing measurements at room (295 K) and cryogenic (4 K) temperatures to characterize the intrinsic piezoelectric acoustic nonlinearity of the Rayleigh and Sezawa guided SAW modes supported in this material stack. The Rayleigh mode exhibits strongly surface-localized strain with coupled longitudinal (in-plane) and vertical (out-of-plane) displacements, whereas the Sezawa mode has a strain-energy profile that extends more deeply into the SiC substrate. Our results show a significant enhancement in four-wave mixing power conversion efficiency for the Rayleigh mode compared to the Sezawa mode, as well as a temperature-dependent enhancement for both modes at 4 K relative to 295 K.

To interpret these results, we apply the standard undepleted four-wave mixing model from Kerr nonlinear optics, which assumes linear loss and a power-independent modal nonlinear coefficient. Within this framework, the measured enhancement in conversion efficiency is consistent with an intrinsic increase in the modal nonlinearity. However, our experimental results exhibit clear departures from the model’s assumptions, revealing nonlinear behavior beyond that captured by a Kerr-type approximation.

\begin{figure*}
\includegraphics[scale = 0.245]{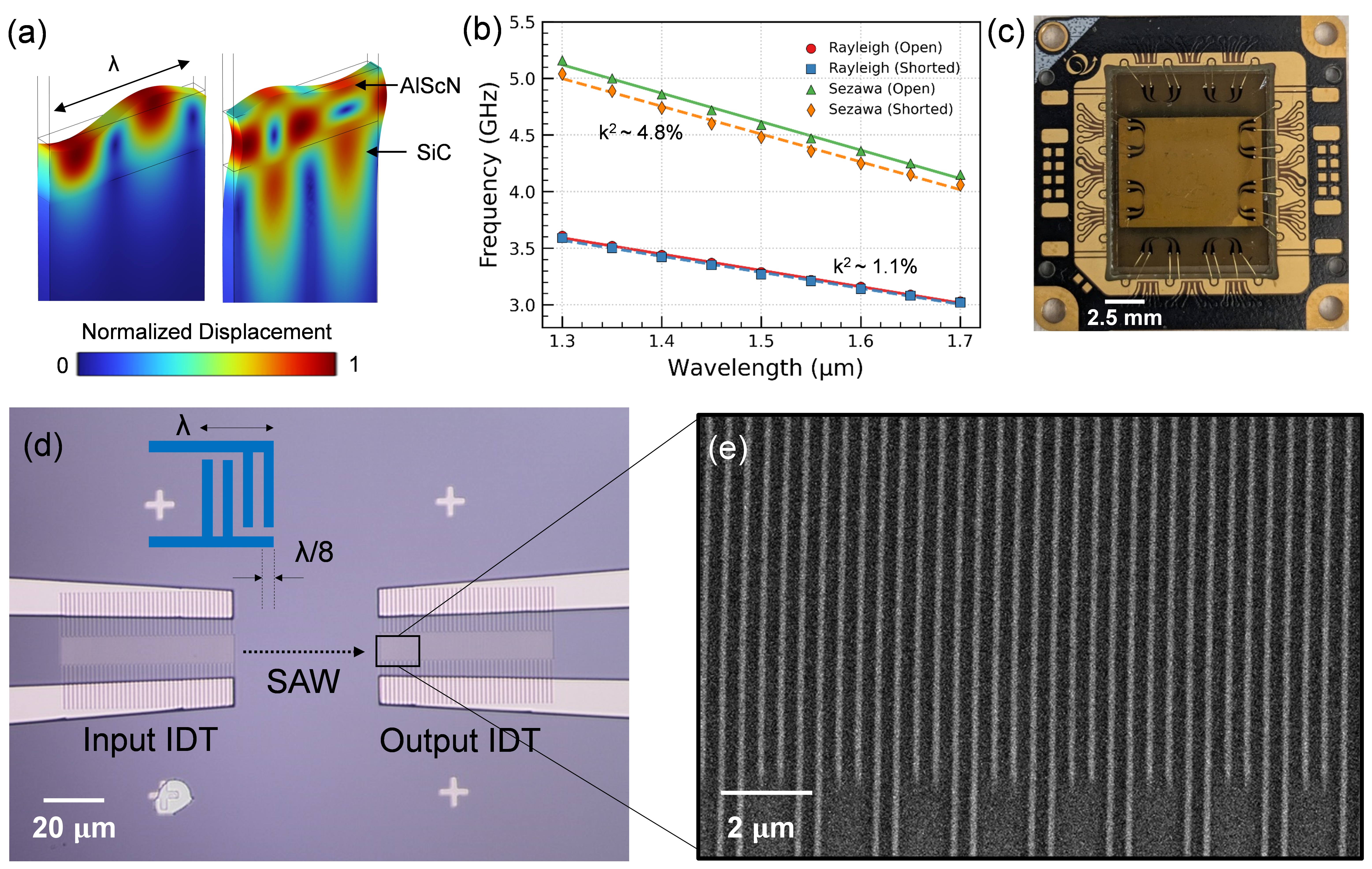}
\caption{\label{fig:1} (a) The modeled normalized displacement field for the Rayleigh mode (left) and Sezawa mode (right) in the 500 nm thick Al$_{0.58}$Sc$_{0.42}$N/SiC material platform for $\lambda$=1.5 \textmu m. (b) Modeled acoustic frequency with electrically open and shorted boundary conditions on the Al$_{0.58}$Sc$_{0.42}$N surface as a function of acoustic wavelength. The modeled $k^2$ values for Rayleigh and Sezawa modes were 1.1\% and 4.8\%, respectively. (c) An image of the fabricated Al$_{0.58}$Sc$_{0.42}$N/SiC chip wirebonded onto a QBoard used for cryogenic measurements. (d) A microscope image of a guided SAW delay line, which consists of an input IDT, a propagation region, and an output IDT. The inset shows a schematic of the split finger IDT design. (e) A scanning electron microscope image of the fabricated IDT electrodes.}
\end{figure*}


The c-axis-oriented Al$_{0.58}$Sc$_{0.42}$N thin film was prepared on a 4H-SiC substrate by reactive pulsed-DC magnetron co-sputtering. The deposition was performed with a heater temperature of 500 $^{\circ}$C and a N$_2$ flow of 20 sccm. Magnetron powers of 875 W and 770 W were applied to the Al and the Sc targets, respectively. To suppress abnormally oriented grains, the AlScN stack was initiated with a 15 nm AlN seed layer followed by a 35 nm compositional gradient layer in which the Sc concentration was linearly increased from 0\% to 42\%.\cite{hackett_s-band_2024,du_near_2024} 
A 450 nm thick layer of Al$_{0.58}$Sc$_{0.42}$N was then grown, such that the total piezoelectric film thickness was 500 nm.  

The interdigital transducers (IDTs) of the piezoelectric acoustic wave delay line were fabricated using e-beam lithography to define the split finger IDT electrodes with an acoustic wavelength, $\lambda$, of 1.5 \textmu m followed by metal evaporation and liftoff. A subsequent photolithography step patterned the bus lines connecting fingers of identical polarity to their contact pads for wirebonding.

The guided SAW delay line device used for four-wave mixing measurements was wirebonded to a Quantum Machines QBoard and measured in a Janis liquid-helium flow cryostat. Scattering parameters (S-parameters) were characterized using a calibrated vector network analyzer (VNA). For four-wave mixing experiments, two RF signal generators provided the pump tones, which were combined and applied to the input IDT. The output spectrum was characterized using a spectrum analyzer with a 1 kHz resolution bandwidth and a 100 Hz video bandwidth. Propagation loss measurements were performed on separate test structures fabricated on a dedicated chip. Those measurements were carried out in a Montana Instruments cryostat using ground–signal–ground (GSG) RF probes. 

\begin{figure}
\includegraphics[scale = 0.13]{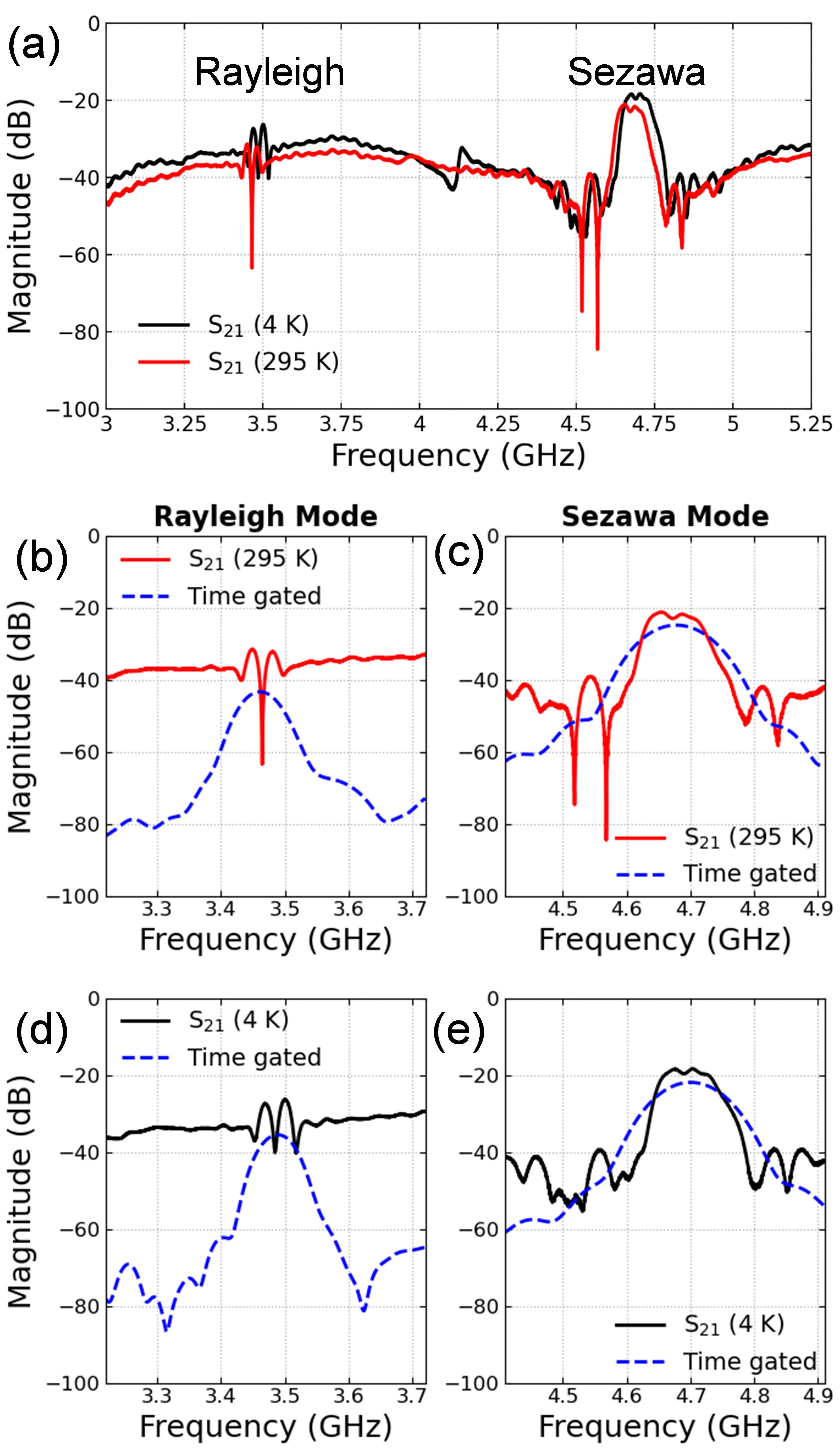}
\caption{\label{fig:2} (a) S$_{21}$ as a function of frequency in the guided SAW delay line measured at 295 K and 4 K, showing Rayleigh and Sezawa modes. (b,c) Measured transmission and time-gated transmission at 295 K for Rayleigh and Sezawa modes. (d,e) Corresponding results at 4 K.}
\end{figure}

Figure \ref{fig:1}(a) shows the computed normalized displacement fields in the Al$_{0.58}$Sc$_{0.42}$N/SiC heterostructure for the Rayleigh and Sezawa modes at an acoustic wavelength of 1.5 \textmu m from a finite element method (FEM) model. The Rayleigh mode has a displacement amplitude that peaks at the free surface and then decreases with increasing depth into the SiC substrate. The majority of the strain energy density is confined within the Al$_{0.58}$Sc$_{0.42}$N layer, with coupled in-plane and out-of-plane motion. In contrast, the Sezawa mode exhibits mixed in-plane and out-of-plane particle motion distributed across the film thickness, with stronger participation near the Al$_{0.58}$Sc$_{0.42}$N/SiC interface and a significantly larger fraction extending into the SiC substrate.

We used FEM simulations to compute the acoustic eigenfrequencies of the Rayleigh and Sezawa modes with electrically open or shorted boundary conditions for a wavelength range of 1.3 \textmu m to 1.7 \textmu m. The resulting mode frequencies as a function of wavelength are shown in FIG. \ref{fig:1}(b). As expected, we find that the Sezawa mode has a higher modeled electromechanical coupling coefficient, $k^2$, of $\approx$4.8\% when compared to the Rayleigh mode, which has a modeled $k^2$ value of $\approx$1.1\%.

Figure \ref{fig:1}(c) shows an optical image of a fabricated Al$_{0.58}$Sc$_{0.42}$N/SiC chip patterned with SAW delay lines wirebonded to a QBoard.  Figure \ref{fig:1}(d) shows a microscope image of a single delay line device that consists of input and output IDTs on the Al$_{0.58}$Sc$_{0.42}$N surface. The IDTs have a wavelength of 1.5 \textmu m, an aperture, $A$, of 10 \textmu m, and are separated by a propagation distance, $L$, of 50 \textmu m. To suppress spurious reflections, we use a split finger IDT design as shown in FIG. \ref{fig:1}(e). 

The guided SAW delay line transmission, S$_{21}$, as a function of frequency is shown in FIG. \ref{fig:2}(a) for 295 K and 4 K. Two transmission peaks appear at $\approx$ 3.5 GHz and $\approx$ 4.7 GHz that we identify as the Rayleigh and Sezawa modes, respectively, from FEM modeling. A significant broadband electromagnetic background is present, which we attribute to uncalibrated sections of the measurement path, which included the cryostat wiring and wirebonds. To isolate the piezoelectric acoustic contribution from the electromagnetic background, we performed a time-gating analysis (FIG. \ref{fig:2}(b--e)). From 295 K to 4 K, both modes exhibit increased transmission and a small resonance frequency shift.

\begin{figure*}
\includegraphics[scale = 0.085]{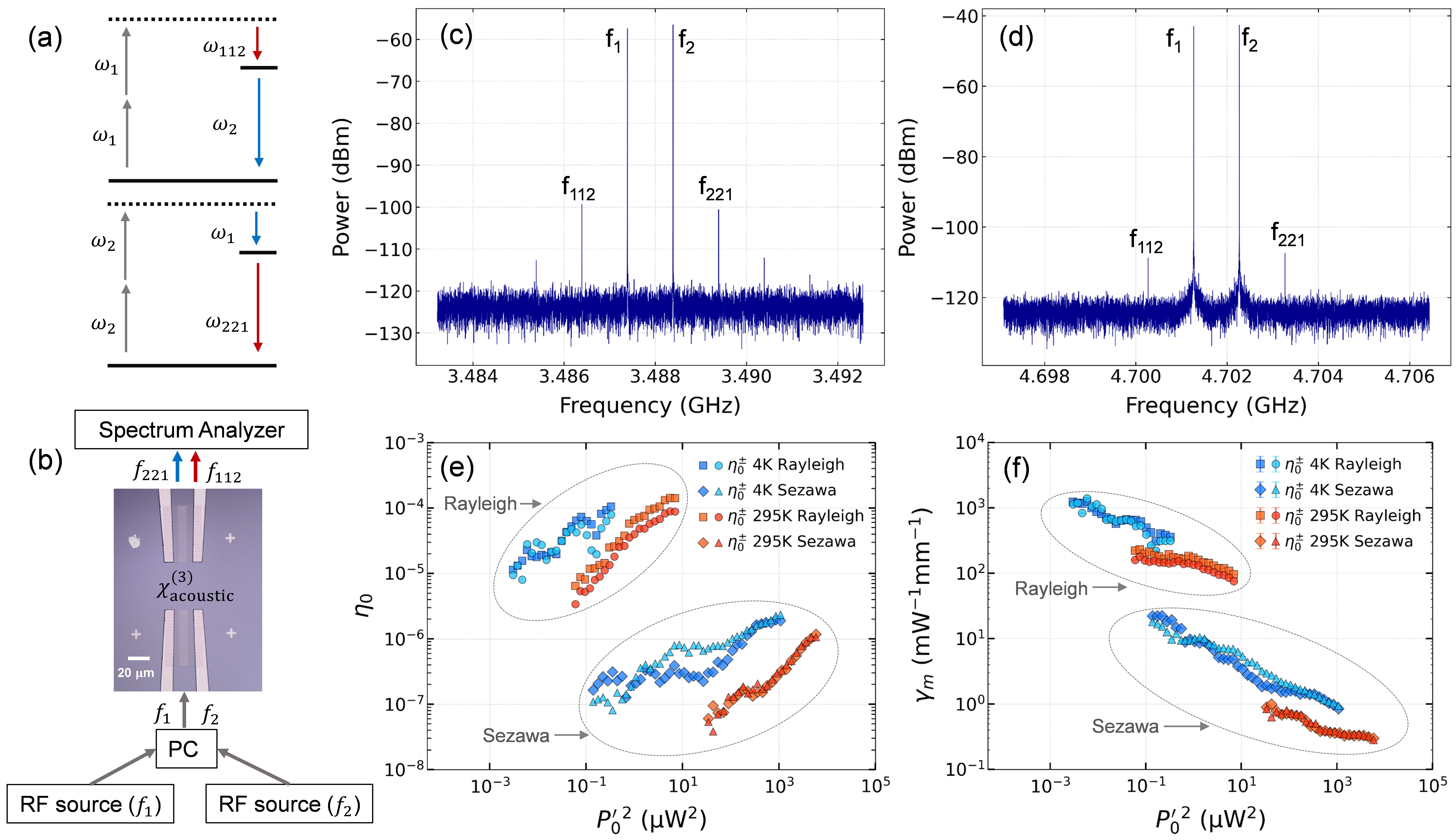}
\caption{\label{fig:3} (a) Energy level diagrams illustrating degenerate four-wave mixing. (b) A schematic of the the four-wave mixing experimental setup. Two RF sources were used to supply two pump tones to the input IDT driven at $f_1$ and $f_2$ frequencies. The output signal was collected by an IDT and characterized using a spectrum analyzer. (c,d) Representative spectra at 4 K for Rayleigh and Sezawa modes, respectively. (e) Extracted power conversion efficiency, $\eta_0$, as a function of $P_0'^2$ at 295 K and 4 K. (f) Extracted modal nonlinear coefficient, $\gamma_m$, as a function of $P_0'^2$.}
\end{figure*}

SAW four-wave mixing arises due to an effective third-order acoustic susceptibility, $\chi^{(3)}_{\text{acoustic}}$, that is expected to depend on factors such as strain localization, strain tensor symmetry, modal overlap with regions of large higher-order elastic and piezoelectric coefficients, nonlinear dielectric response, and interfacial effects. When two phononic pump tones at angular frequencies $\omega_1$ and $\omega_2$ are launched into the delay line, degenerate four-wave mixing generates sidebands at $\omega_{112}=2\omega_1-\omega_2$ and $\omega_{221}=2\omega_2-\omega_1$, as shown in FIG. \ref{fig:3}(a). The four-wave mixing measurement method utilized here is shown in FIG \ref{fig:3}(b). Two RF signal generators supplied pump tones at frequencies $f_1$ and $f_2=f_1+\Delta f$ where $\Delta f$ = 1 MHz such that both pump tones remained within the IDT passbands for the Rayleigh and Sezawa modes. The pump tones were combined and applied to the input IDT. As the acoustic waves propagated through the Al$_{0.58}$Sc$_{0.42}$N/SiC heterostructure, the third-order acoustic nonlinearity generated four-wave mixing sidebands, which were transduced back into electrical signals at the output IDT and recorded using a spectrum analyzer. For the first-order sidebands considered here, phase mismatch is negligible because the interaction region is short with an effective length, $L_{\text{eff}}$, of approximately 50 \textmu m and the acoustic dispersion over the relevant frequency span is weak. Using FEM-computed acoustic phase velocities to evaluate the wavevector mismatch, $\Delta k$, we find $|\Delta k|L_{\mathrm{eff}}\ll 1$.  

Figure \ref{fig:3}(c) shows the measured four-wave mixing spectrum for the Rayleigh mode at 4 K with two pump tones centered at $\approx$3.488 GHz. The RF signal generator power was set to -10 dBm. Two distinct sidebands labeled as $f_{112}$ and $f_{221}$ appear at a $\Delta f$ spacing away from $f_1$ and $f_2$, respectively. Figure \ref{fig:3}(d) shows the corresponding spectrum for the Sezawa mode with the pump tones centered around $\approx$4.702 GHz. For clarity, we label the primary pump tones as $n=0_-$ and $0_+$ and the successive sidebands as $n = \pm 1, \pm2, ...$.

The power conversion efficiency (PCE), $\eta_{0\pm}$, for degenerate four-wave mixing is given by
\begin{equation}
    \eta_{0\pm}=\bigg|\frac{P_{1\mp}}{P_{0\pm}}\bigg|=\Gamma P_{0\mp}'^2
    \label{eq:1}
\end{equation}
where $P_{1\mp}$ is the measured sideband power, $P_{0\pm}$ is the measured pump tone power, $P_{0\mp}'$ is the input acoustic pump power at the beginning of the guided SAW waveguide, and $\Gamma$ is the four-wave mixing coefficient. $P_{0\mp}'$ was determined by assuming symmetry of the experimental setup. The total loss, after subtracting propagation loss and the measured power combiner loss, was evenly divided between input and output losses which include contributions from cryostat cabling, wirebonds, and IDT transduction. The calculated input loss was then subtracted from input power to determine $P_{0\mp}'$. The propagation losses were independently measured as discussed later.

Figure \ref{fig:3}(e) shows the extracted PCE for the Rayleigh and Sezawa modes at 295 K and 4 K on a log-log scale as a function of $P_0'^2$. In the low power regime, we include only data for which the sideband powers are at least 4 dB above the spectral noise floor. At higher pump powers, we exclude data once the PCE clearly departs from the expected $P_0'^2$ scaling. At 295 K, the maximum PCE for the Sezawa mode was 1.5$\times$10$^{-4}$\% at an acoustic input power of 78.2 \textmu W whereas a significantly larger maximum PCE of 1.7$\times$10$^{-2}$\% was achieved for the Rayleigh mode at an acoustic input power of 2.63 \textmu W. We calculated the four-wave mixing coefficient, $\Gamma$, from Eq. (\ref{eq:1}) by fitting the slope of measured PCE with respect to the input power. At 295 K, $\Gamma=(1.9\times 10^{-4}\pm 1.3 \times 10^{-5})$ mW$^{-2}$ was obtained for the Sezawa mode, whereas a significantly larger value of $\Gamma=(50.4\pm 3.2)$ mW$^{-2}$ was extracted for the Rayleigh mode. Furthermore, we observed an enhancement in the extracted four-wave mixing coefficient at 4 K as compared to 295 K where values of $\Gamma=(3.9\times10^{-3}\pm 1.6\times10^{-4})$ mW$^{-2}$ and $\Gamma=(813\pm 48)$ mW$^{-2}$ were obtained for Sezawa and Rayleigh modes at 4 K, respectively. The errors in the listed $\Gamma$ values were obtained from the accuracy of the fits.

The modal nonlinear coefficient, $\gamma_m$, can then be calculated from the extracted $\Gamma$ according to
\begin{equation}
    \gamma_m = \frac{\sqrt{\Gamma}}{L_{\text{eff}}}
    \label{eq:2}
\end{equation}
where $L_{\text{eff}}$ is the effective nonlinear interaction length given by
\begin{equation}
    L_{\text{eff}} = \bigg(\frac{1-e^{-\alpha L}}{\alpha}\bigg)
    \label{eq:3}
\end{equation}
where $L$ is the physical length of the guided SAW waveguide and $\alpha$ is the acoustic propagation loss. To estimate the acoustic propagation loss in the Al$_{0.58}$Sc$_{0.42}$N/SiC heterostructure, we fabricated multiple waveguides with identical split finger IDTs and varying propagation lengths ranging from 50 \textmu m to 500 \textmu m with a 50 \textmu m aperture. The 50 \textmu m aperture was used to suppress diffraction losses at longer propagation lengths. We extracted $\alpha$ from peak transmission as a function of the propagation distance. At 295 K, the extracted propagation losses were (106 $\pm$ 19) dB/cm for the Rayleigh mode and (96 $\pm$ 8) dB/cm for the Sezawa mode. After cooling to 4 K, the propagation losses were reduced to (8 $\pm$ 40) dB/cm and (24 $\pm$ 8) dB/cm for the Rayleigh and Sezawa modes, respectively. The substantially larger relative uncertainty for the Rayleigh mode at 4 K arises from the formation of a weak acoustic cavity due to reflections in the device, despite the use of a split finger IDT design. 

By using the values for $\alpha$ and $\Gamma$ in Eq. (\ref{eq:2}) and Eq. (\ref{eq:3}), we determined the modal nonlinear coefficients. For the Rayleigh mode, $\gamma_m =($151 $\pm$ 9) mW$^{-1}$mm$^{-1}$ at 295 K and (573 $\pm$ 22) mW$^{-1}$mm$^{-1}$ at 4 K. For the Sezawa mode, on the other hand, at 295 K, $\gamma_m =$(0.3 $\pm$ 0.04) mW$^{-1}$mm$^{-1}$ with an enhancement to (1.3 $\pm$ 0.2) mW$^{-1}$mm$^{-1}$ at 4 K. The errors in the reported $\gamma_m$ originate from the relative errors in $\Gamma$ and $\alpha$. For the undepleted, Kerr-type model described by Eqs. (\ref{eq:1})--(\ref{eq:2}), one would expect the modal nonlinear coefficient, $\gamma_m$, to be independent of input power. However, by investigating the effective $\gamma_m$ extracted at each input power (FIG. \ref{fig:3}(f)), we observe clear power-dependent behavior, indicating departures from the simple Kerr-type description at higher acoustic input power. 

At higher input power, nonlinear processes beyond degenerate four-wave mixing may divert acoustic power from the measured four-wave mixing sidebands. These additional frequency components are strongly attenuated by the output IDT and therefore are not able to be captured in measured spectra. At large pump amplitudes, higher odd-order nonlinearities can generate cross-Kerr like sidebands at the same offset frequencies as those produced by the third-order process. These are indistinguishable from $\chi^{(3)}_{\text{acoustic}}$ driven sidebands in our measurements, but can exhibit substantially different pump-power scaling and relative sideband phases. An amplitude-dependent reduction of $\chi^{(3)}_{\text{acoustic}}$, which could arise from anharmonic elastic saturation, nonlinear dissipation, or mode-profile distortion at high strain, also cannot be excluded with only the data presented here.

To ensure that we only recorded the nonlinear properties of the guided SAW waveguide, we performed a control four-wave mixing measurement through the wirebonded QBoard bypassing our fabricated phononic waveguide. Under the spectrum analyzer settings used for this work, we observed no sideband generation for the input powers reported here, confirming that the measured four-wave mixing response is from the guided SAW waveguide.

The undepleted, phase-matched four-wave mixing model (Eqs. (\ref{eq:1}-\ref{eq:2})), assumes single-pass propagation and neglects pump build-up from IDT reflections. For a 50 \textmu m delay length with low $\alpha$, even small IDT reflectivity can form a weak acoustic cavity. We quantify this with a resonant build-up factor:
\begin{equation}
    B=\frac{1}{(1-Re^{-\alpha L})^2}
    \label{eq:4}
\end{equation}
where $R$ is the IDT reflectivity. The intracavity pump power is then $P^{cavity}_{0 \mp}=BP'_{0\mp}$.  Equation (4) assumes excitation on resonance and therefore provides an upper bound on the intracavity pump power enhancement and correspondingly on the cavity-induced modification of the modal nonlinearity. Here, we have treated the guided SAW waveguide as a phononic analogue of an optical Fabry-Perot cavity with internal resonance power enhancement with identical mirrors.\cite{boyd_nonlinear_2008} 

IDT reflectivity was extracted from passband transmission ripples according to $R = \frac{1-\sqrt{1-V^2}}{V}e^{\alpha L}$, where $V = \frac{\sum^{n}_{i=1}(\frac{T_{max}-T_{min}}{T_{max}+T_{min}})}{n}$ is the fringe visibility calculated for $n$ ripples where each ripple has a maximum and minimum transmission denoted by $T_{max}$ and $T_{min}$, respectively.\cite{pedrotti_introduction_2017} We obtained reflectivities of 0.136 and 0.098 at 295 K for the Rayleigh and Sezawa modes, respectively. At 4 K, the reflectivity values increased to 0.3 for the Rayleigh mode and 0.135 for the Sezawa mode. Even with a substantial uncertainty in the propagation loss for Rayleigh mode at 4 K, the extracted reflectivity only changes by 10\% because the propagation distance is short and therefore $e^{\alpha L}$ is close to unity. The enhanced reflectivity that we extract at 4 K could be due to reduced resistive losses in the Al IDTs at cryogenic temperatures producing altered boundary conditions at the AlScN/Al interfaces. 

Including pump build-up, 
\begin{equation}
    \gamma_m = \frac{\sqrt{\Gamma}}{BL_{\text{eff}}}.
    \label{eq:5}
\end{equation}
 According to Eq. \ref{eq:5}, the Rayleigh mode four-wave mixing modal nonlinearity is ($117\pm7$) mW$^{-1}$mm$^{-1}$ at 295 K and ($283\pm11$) mW$^{-1}$mm$^{-1}$ at 4 K. The values for the Sezawa mode are ($0.24 \pm0.03$)  mW$^{-1}$mm$^{-1}$ and ($0.96\pm0.1$) mW$^{-1}$mm$^{-1}$ at 295 K and 4 K, respectively, with uncertainties originating from the four wave mixing coefficient, $\Gamma$, and propagation loss, $\alpha$. We therefore find that the cavity-enhanced pump power cannot account for the large $\gamma_m$ observed for the Rayleigh mode over the Sezawa mode and for 4 K compared to 295 K for both modes.


Our results suggest that the Rayleigh mode in the Al$_{0.58}$Sc$_{0.42}$N/SiC heterostructure supports an intrinsically large four-wave mixing modal nonlinear coefficient while the Sezawa mode may be more well-suited for applications where it is desirable to suppress the third-order nonlinearity. The spatial distributions of displacement fields and strain energy density play a critical role in determining nonlinear response. Incorporating lateral confinement into this material system via AlScN etching, which has already been demonstrated,\cite{deng_monolithic_2025} could provide a tool for engineering nonlinear overlap. More broadly, the dependence of the intrinsic modal nonlinearity on waveguide width and effective mode area remains largely experimentally unexplored for phononic systems. In nonlinear optics, the modal nonlinear coefficients generally scale inversely with effective mode area, but the relationship is often more complex due to effects such as field redistribution and dispersion.

The observed enhancement of phononic four-wave mixing at cryogenic temperatures does have direct analogues in nonlinear photonics, where Kerr-type nonlinear processes often exhibit increased efficiency at low temperatures due to reduced linear loss, suppression of thermally activated dissipation channels, and reduced nonlinear dissipation channels. Future temperature-dependent measurements in phononic systems can potentially assess the relative roles of intrinsic anharmonicity, nonlinear loss, and energy redistribution in cryogenic nonlinear acoustic interactions. It is also possible that the intrinsic phononic nonlinearity itself increases at cryogenic temperatures due to temperature-dependent higher-order elastic, piezoelectric, or dielectric coefficients, as well as temperature-dependent interfacial modifications.

Electron-mediated phononic nonlinearities also offer a route to enhance four-wave mixing.\cite{hackett_giant_2024} An Al$_{0.58}$Sc$_{0.42}$N/SiC heterostructure has already been successfully integrated with a semiconductor thin film,\cite{hackett_s-band_2024} providing a clear path to study this method for enhancing phononic nonlinearities for the Rayleigh and Sezawa modes. Extending such approaches to cryogenically compatible and quantum platforms would require alternate semiconductor films with low carrier freeze-out, such as graphene.


In this work, we have demonstrated phononic four-wave mixing in an Al$_{0.58}$Sc$_{0.42}$N/SiC heterostructure, investigating both Rayleigh and Sezawa guided SAW modes at room temperature (295 K) and cryogenic temperature (4 K). We found that the Rayleigh mode exhibits a substantially larger modal nonlinear coefficient than the Sezawa mode, with an enhancement of approximately 450X at 4 K and 500X at 295 K. In addition, both modes show an enhancement of approximately 4X in the four-wave mixing modal nonlinearity at 4 K relative to 295 K. These results highlight the strongly modal-dependent and temperature-dependent nature of phononic nonlinear interactions in guided SAW systems.

\section{\label{Acknowledgments} Acknowledgments}
 This work is supported by the Laboratory Directed Research and Development program at Sandia National Laboratories, a multimission laboratory managed and operated by National Technology and Engineering Solutions of Sandia LLC, a wholly owned subsidiary of Honeywell International Inc. for the US Department of Energy’s National Nuclear Security Administration under contract DE-NA0003525. This work was performed, in part, at the Center for Integrated Nanotechnologies, an Office of Science User Facility, operated for the US Department of Energy Office of Science. This paper describes objective technical results and analysis. Any subjective views or opinions that might be expressed in the paper do not necessarily represent the views of the US Department of Energy or the US Government.


\section{\label{Data Availability} Data Availability}
The data that support the findings of this study are available from the corresponding author upon reasonable request.

\section{\label{References} References}

\bibliography{ref_final_v6}

\end{document}